\documentclass[10pt,letterpaper]{article}
\usepackage[top=0.85in,left=2.75in,footskip=0.75in]{geometry}
\usepackage{authblk}

\usepackage{amsmath,amssymb}

\usepackage{changepage}

\usepackage[utf8x]{inputenc}

\usepackage{textcomp,marvosym}

\usepackage{cite}

\usepackage{nameref,hyperref}

\usepackage{romannum}

\usepackage{microtype}

\usepackage[table]{xcolor}
\usepackage[normalem]{ulem}

\usepackage{array}
\usepackage{siunitx}
\usepackage{tikz}
\usepackage[english]{babel}
\usepackage[autostyle]{csquotes}
\usepackage{subfig}
\usepackage[utf8x]{inputenc}

\newcolumntype{+}{!{\vrule width 2pt}}

\newlength\savedwidth

\raggedright
\usepackage[aboveskip=1pt,labelfont=bf,labelsep=period,justification=raggedright,singlelinecheck=off]{caption}

\makeatletter
\renewcommand{\@biblabel}[1]{\quad#1.}
\makeatother

\usepackage{lastpage,fancyhdr,graphicx}
\usepackage{epstopdf}

\fancyheadoffset[L]{2.25in}
\fancyfootoffset[L]{2.25in}
\begin{document}
\vspace*{0.2in}

\begin{flushleft}
{\Large\textbf\newline{High participation ratio genes in the interaction network structure} }
\newline

\ Nastaran Allahyari\textsuperscript{1},\ Nima Abedpour\textsuperscript{2},\ Ali Hosseiny\textsuperscript{1},\ G. R. Jafari\textsuperscript{1,3*}

\bigskip
\textbf{1} Department of Physics, Shahid Beheshti University, Evin, Tehran, Iran, \textbf{2} Department of Translational Genomics, University of Cologne, Weyertal, Cologne, Germany, \textbf{3} Institute of Information Technology and Data Science, Irkutsk National Research Technical University, Lermontova, Irkutsk, Russia  
\\
\bigskip

* g\_jafari@sbu.ac.ir (G. R. J)

\end{flushleft}

\section*{Abstract}

Genes have specific functional roles, however, since they are dependent on each other, they can play a structural role within a network structure of their interactions. In this study, we analyze the structure of the gene interaction network and detect the most contributing genes through the random matrix theory. Specifically, we compare the interaction network of essential and nonessential genes of the yeast Saccharomyces cerevisiae. Most remarkably, this well-established combined framework by measuring the node participation ratio $(NPR)$ index helps detect important genes, which control the insightful structural patterns in the underlying networks. Results indicate that the essential genes have higher values of $NPR$ rather than the nonessential ones which means that they have the most contribution to the network structure. It is worth mentioning that among all essential genes, the $NPR$ value of 5 significant ones is considerably higher than the other essential genes, and also the same is for 15 significant nonessential genes compared to the others. Thus, the significant essential genes strongly manage their network structure, while the significant nonessential genes, besides their global contributions, have weak effects on their network structure. Most strikingly, these genes existing in a limited number of structural patterns are responsible for the specific bioprocesses which are the signature of their networks.

\section*{Introduction}

The abundance of available data at the genomic level and the considerable progress in high-tech methods in data collection, instead of gene-based biology, has resulted in network biology \cite{Battiston, Barabasi1}. When a gene interaction network is analyzed, the interaction between two genes is interestingly affected by the other interconnections they participate within \cite{Szappanos, Cordell, Tong, Marbach}. Indeed, the dependency of interactions through a network can create a structure where the emergence of collective behavior may occur. Thus, uncovering the mechanism leading to collective behavior in networks is considered a significant subject in biology. To this aim, we explore the weighted, signed, and undirected networks of genetic interaction profile similarities of Saccharomyces cerevisiae. In this paper, to obtain a deeper understanding of the role and importance of interaction patterns in our networks, a time-efficient and cost-effective mathematical tool, random matrix theory $(RMT)$ is applied. Through this study, the crucial questions are: Do the genetic interaction networks display specific evidence to deviate from randomness and clarify heterogeneous structures \cite{Orsini}? How to detect genes that play a leading role in forming heterogeneities in the network? This work, by $RMT$ techniques, not only deals with the importance of structural patterns in the networks but also helps in detecting genes, which are significant for the networks' identity.

There is a long history in which the non-randomness of gene interaction networks has been studied by considering motifs as the simplest units of network structure \cite{Shen, Sanz1, Sanz2, Lee, Allahyari, Kargaran2}. Here, we explore the genetic interaction networks compared to the shuffled ones in the context of $RMT$. This method was initially introduced by Eugene Wigner \cite{new_Wigner_1} to study the spectrum of the nuclei of heavy atoms where a large number of nucleons are interacting. It has recently demonstrated its outstanding success in analyzing complex systems in various fields such as quantum chaos \cite{Papenbrock}, mobility \cite{Louail}, ecological \cite{Stone, Grilli, Allesina}, and financial \cite{Urama, Saeedian, Jamali} networks. Moreover, it has provided a valuable framework to study complex biological networks. One application of $RMT$ in biological networks is to study the spectral fluctuation of interaction matrices indicating the Gaussian orthogonal ensemble $(GOE)$\cite{Lou-7, Luo} of $RMT$. Also, it helps the overall functioning of altered complex molecules and predicting the connections among the molecules as one of the main goals in modern biology \cite{Rai-4}. Moreover, through $RMT$, the cellular roles of some unknown genes are predicted in terms of performance \cite{Gibson}. Besides, due to inadequacy in the theoretical study to analyze gene interaction networks in cancer cells, $RMT$ plays a crucial role to predict the behavior of a large complex interaction network from the behavior of a finite size network \cite{Kikkawa-1}. Furthermore, as another application, information about the structure and dynamics of a network, through the localization properties of the gene network is achieved via the eigenvalues of its adjacency matrix \cite{Jalan2, Rai-8, Rai-2, Kikkawa-333}. Additionally, $RMT$ has helped solve the challenge of revealing all of the gene-gene interactions that are large in number \cite{Frost}. As a recent application, we can refer to \cite{Aparicio-5}, where eigenvalues and eigenvectors denoise single-cell data, which provides the opportunity to identify new cellular states. Margaliot et al. have developed a new theoretical framework using tools from $RMT$ to analyze the steady-state production rate affected by many stochastic  \enquote{local} factors \cite{Margaliot}, as well.

Accordingly, we are interested in determining the heterogeneities of our gene interaction networks and identifying the genes that play a significant role in this structure. In this work, we explore the genome of Saccharomyces cerevisiae, as a whole system to understand the involved complexity. Here, we focus on the gene interaction similarity networks of about 1000 essential genes and 4500 nonessential genes.  Costanzo and his colleagues have produced the data. Thus, we lead our analysis in two groups of genes, namely, essential and nonessential. By considering the threshold taken in their study \cite{Costanzo} to plot the networks, the essential gene similarity network is more densely connected compared to the corresponding nonessential network. Thus, it can be deduced that essential genes play the role of hubs in the global interaction network  \cite{Blomenand, Winzeler, Giaever, Wang}. Moreover, functional relationships in the gene similarity network for essential genes are stronger and their power of function prediction provided higher accuracy \cite{Costanzo2, Baryshnikova, Deshpande, Li}.

The rest of the paper is organized as follows. First, we demonstrate through the structural parameters namely degree, clustering coefficient, and assortativity coefficient that the structure of both essential and nonessential gene networks is far from randomization. Second, we explore the genetic interaction networks via eigenvalue distributions in comparison with their corresponding shuffled versions to analyze the existence of heterogeneous structures. Since in random networks, the spectrum of eigenvalues fits a semi-circle, we discuss that the deviation from such behavior is a sign of structure in the pairwise interactions \cite{Levine}. Next, we study the nearest neighbor spacing distribution for both networks to understand the universality and to check if we are allowed to apply $RMT$ in case they follow the $GOE$ of $RMT$. Then, we investigate the localization properties based on the participation of all genes in each eigenvector via calculating the inverse participation ratio in the gene networks. We discuss that the analysis of the nodes with the top contributions in the localized eigenvectors demonstrates their importance in the structural patterns of the networks. Finally, as a novel index in $RMT$, we measure node participation ratio $(NPR)$ of the important genes in all eigenvectors to separate significant genes which have higher $NPR$ than the boundary of the shuffled networks. Interestingly, the selected significant genes leading to the structural patterns of each network have the most responsibility for the specific bioprocesses which are merely appointed to the same network.

\section*{Materials and methods}

\subsection*{Data description}

{\bf Essential and nonessential genes.} Saccharomyces cerevisiae is known as a helpful yeast since nearly all bioprocesses in eukaryotes can exist in it \cite{Parapouli}. This paper, to achieve a deeper understanding of the complexity of its genome, analyzes its interaction similarity networks. Costanzo and his colleagues have provided and published the data \cite{Costanzo} as three gene interaction similarity matrices, for essential genes, nonessential genes, and the combination of them as the global form \cite{Baryshnikova, Li}. What follows is the specific characteristics that based on them, the genes are categorized into two groups of essential and nonessential ones. The type of mutation generating these mutants for essential and nonessential genes are temperature-sensitive and deletion mutations, respectively. Besides, essential genes reveal a stronger functional connection and higher-accuracy gene function predictions for biological processes. Moreover, the biological processes specifically annotated by the essential genes are \enquote{cell polarity and morphogenesis}, \enquote{ protein degradation}, and \enquote{ribosomal $RNA$ $(rRNA)$ and non-coding $RNA$ $(ncRNA)$ processing}. While, the bioprocesses such as \enquote{nuclear-cytoplasmic transport}, \enquote{ribosome biogenesis}, \enquote{peroxisome}, \enquote{metabolism and fatty acid biosynthesis}, \enquote{respiration, oxidative phosphorylation, mitochondrial targeting}, \enquote{$MVB$ sorting and $pH-$dependent signaling}, and \enquote{$tRNA$ wobble modification} are identified by nonessential genes. In addition, there are six common bioprocesses between both essential and nonessential gene interaction networks that are presented in Table\ref{table0} beside specific bioprocesses in each network \cite{Costanzo}.

\begin{table}[!h]
	\centering
	\fontsize{7}{11}
	\caption{Classification of bioprocesses in essential and nonessential gene networks.}
	\label{table0}
	\centering
	\begin{tabular}{|c|c|c|c}
		\hline
		\rowcolor{lightgray}
		\multicolumn{3} {| c |}{$\bf{GO \ biological \ process \ in \ : }$}\\
		\hline
		\rowcolor{lightgray}
		$\bf{Essential \ network}$ & $\bf{Nonessential \ network}$ & $\bf{Both   \ networks}$ \\
		\hline
		\shortstack{ \\$cell \ polarity$ \ \\ $and \ morphogenesis$}  & \shortstack{ \\$respiration,$ \\$oxidative \ phosphorylation,$ \\ $mitochondrial \ targeting$} & \shortstack{ \\$mitosis$ \\$ \ and \ chromosome \ segregation$} \\
		\hline		
		\shortstack{ \\$protein \ degradation$ \\$ \ / \ turnover$} & \shortstack{ \\$MVB \ sorting$ \\$\ and \ pH-dependent \ signaling$} & $vesicle \ traffic$ \\
		\hline
		\shortstack{ \\$rRNA$ \\$\ and \ ncRNA \ processing$} & $tRNA \ wobble \ modification$ & \shortstack{ \\$transcription$ \\$ \ and \ chromatin \ organization$} \\
		\hline
		$-$ & $ribosome \ biogenesis$ & \shortstack{ \\$glycosylation,$ \\$ \ protein \ folding/targeting,$ \\$ \ cell \ wall \ biosynthesis$} \\
		\hline
		$-$ & $peroxisome$ & $DNA \ replication \ and \ repair$ \\
		\hline
		$-$ & $nuclear-cytoplasmic \ transport$ & $mRNA \ processing$ \\
		\hline
		$-$ & \shortstack{ \\$metabolism$ \\$ \ and \ fatty \ acid \ biosynthesis$} & $-$ \\
		\hline
	\end{tabular}
\end{table}

{\bf Gene interaction matrices.} The data analyzed during the current study comprehensively covers  $\sim 90\%$ of all yeast genes and is publicly available at \url{http://boonelab.ccbr.utoronto.ca/supplement/costanzo2016/}. We have worked with data file $S3$ titled \enquote{Genetic interaction profile similarity matrices}. The steps taken to create this data are as follows:

{\begin{enumerate}
		\item Based on the growth rate of a colony size including two specific mutated genes, the genetic interaction score (epsilon)  of them has been obtained.
		\item Each mutated gene is crossed to an ordered array of other mutated genes. Then, a genetic interaction profile for that gene is constructed.
		\item By calculating the Pearson correlation coefficient, the similarity between each two genetic interaction profiles has been produced.
	\end{enumerate}

Furthermore, the preprocessing procedure on data to reach the gene interaction networks is algorithmically plotted in Fig\ref{fig1}. To be more specific, in the genetic interaction matrix, the positive value for every two genes means how much their profiles are functionally similar to each other, and vice versa. Moreover, the larger the positive (negative) interaction value of the two genes, the more similarity (dissimilarity) between them. Also, elements with zero value indicate that there is no relation between those two genes functionally. Besides, the size of essential, nonessential, and global matrices are $\sim$$1000$, $4500$, and $5500$, respectively. Thus, all our matrices here are undirected, signed, and weighted adjacency matrices. To have a vision of the construction of the networks, in Fig\ref{fig1} they are graphed by using the spring-embedded layout algorithm in the Cytoscape software \cite{Shannon}. Only here, the same threshold taken previously by Costanzo et al. \cite{Costanzo} has been used, i.e., if the interaction between two genes is more than $0.2$, they are connected, otherwise, there is no connection between them. We then investigate the structural properties of these networks.

\begin{figure}[h]
	\centering
	{\includegraphics[height=13.cm,width=1.\linewidth]{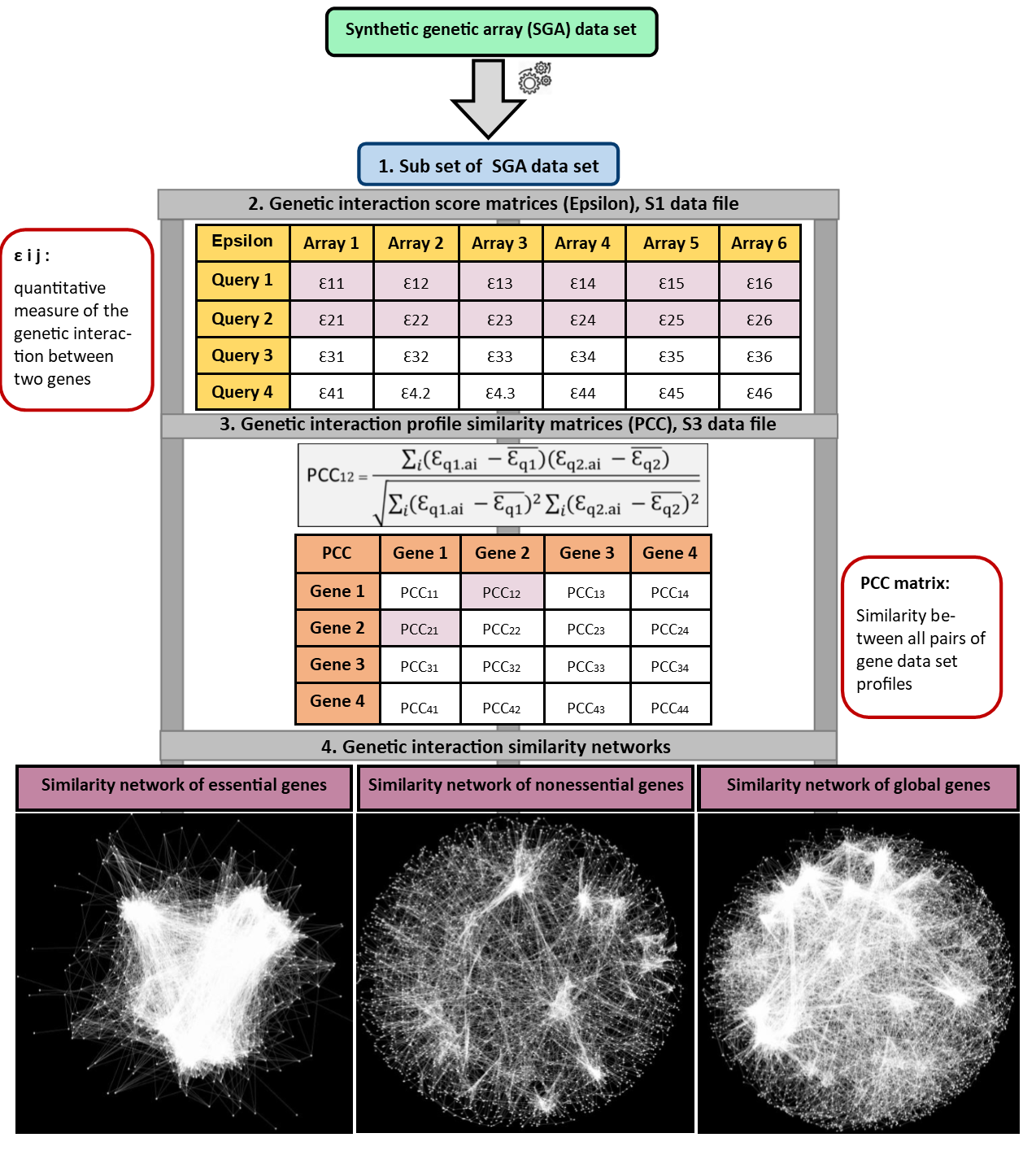}}
	\caption{\bf Graphical abstract for the procedure of obtaining the genetic interaction similarity matrices.}
	\label{fig1}
\end{figure}

\subsection*{Structural measures}

{\bf Degree $k$.} Several statistical indicators are proposed to comprehend specific features of the network \cite{Boccaletti}. The most primary structural measure of a network is the degree of a node $k_{i}$, which is defined in a weighted and signed network as the sum over the absolute of links' weights of a node ($k_{i} = \sum_{j} |w_{ij}|$). Besides, mean degree $\langle k \rangle$ shows the degree of nodes in a network on average.

{\bf Assortativity Coefficient.} To understand that high-degree components tend to be connected with similar (different) counterparts, the assortativity (disassortativity) coefficient is calculated. If similar (different) counterparts are connected, then the sign of this coefficient is positive (negative).
\\
{\bf Clustering Coefficient $CC$.} Another significant parameter is the clustering coefficient $CC$ of the network. The $CC$ of a network characterizes the overall tendency of nodes to form clusters or groups. It is defined as the ratio of the number of closed triplets it has by the number of open triplets it has. These closed triplets are complete subgraphs in the network, which are also known as cliques-3. As proposed in \cite{Costantini}, in a signed, undirected, and weighted network, the clustering coefficient of a node $CC_{i}$ and of the network $CC$ can be calculated, respectively as

\begin{equation}
CC_{i} = \frac{\sum_{j,k} (w_{ij}w_{jk}w_{ki})}{\sum_{j\neq k}|w_{ij}w_{ik}|},
\label{eq1}
\end{equation}

\begin{equation}
CC = \frac{\sum_{i} \sum_{j,k} (w_{ij}w_{jk}w_{ki})}{\sum_{i} \sum_{j\neq k}|w_{ij}w_{ik}| + \sum_{i,j}w_{ij}^2},
\label{eq2}
\end{equation}
where $w$ stands for the weight of the link between every two genes in the network.

In the following, to obtain a deeper understanding of the complexity and importance of structural patterns in our networks, the use of random matrix theory 
$(RMT)$ as a time-efficient and cost-effective mathematical tool is explained. Moreover, $RMT$ as a powerful and frequently used method can provide an approach to identify significant nodes implicated in each network structure. Therefore, the global properties such as spectral properties, local properties such as eigenvalue fluctuations, and detection of significant genes appearing in the localized eigenvectors are studied. 

\subsection*{Spectral techniques}

{\bf Spectral distribution of eigenvalues $P(\lambda)$.}} Our approach to studying the genetic interaction similarity networks is $RMT$ that provides a framework to go beyond the assumption that pair interactions are independent of each other in networks. When a large number of elements interact, and a random relation mainly drives the interactions, and the distribution of the components of the interaction matrix is Gaussian, then the $P(\lambda)$ has a semi-circle shape. Specifically, in a random network with zero mean value of its eigenvalues, all eigenvalues are collected around zero. But in a random network with a nonzero mean value of its eigenvalues, there is an eigenvalue just out of the bulk. Similarly, in the shuffled network, the distribution of eigenvalues has the same form as the bulk \cite{Rai-4}. In the shuffling process, two elements of the matrix are chosen randomly and are exchanged. As a result, neither the value of links nor their distribution is modified, but the correlation between links dissolves. Therefore, when the process is complete, the shuffled network has random connections with no specific structure \cite{Leskovec}.  Despite the shuffled networks, the original networks have a large class of eigenvalues out of the bulk. To define the boundary of the bulk, we shuffle the matrix, then put the largest eigenvalue aside. This experiment is repeated $100$ times. Then, the mean value of the borders and the standard deviations of them are worked out. Accordingly, all eigenvalues at least three standard deviations away from the border are considered as the eigenvalues outside of the bulk.

{\bf Nearest neighbor spacing distribution $P(s)$.} There are three categories of $RMT$ depending on their symmetries, the Gaussian orthogonal ensembles $(GOE)$, the Gaussian unitary ensembles $(GUE)$, and the Gaussian symplectic ensembles $(GSE)$. Here, as we study the case that the gene interaction matrices are real symmetric, thus, $GOE$ of $RMT$ is analyzed. In $RMT$, besides the $P(\lambda)$ as a global property, eigenvalue fluctuation which is related to the symmetry of the system as local properties is considered. To this aim, one of the most powerful techniques in $RMT$, i.e., the nearest neighbor spacing distribution $P(s)$ is studied. In the large matrix size, the $P(s)$ becomes a universal function. Here,  “universal” denotes that the distribution is independent of any detail of the systems and is only concerned with its symmetry. In the $GOE$ case,

\begin{equation}
P(s) = \frac{\pi s}{2} \exp (\frac{-\pi s^2}{4}),
\label{eq3}
\end{equation}

where $s$ is the spacing between two adjacent eigenvalues normalized by the mean eigenvalue spacing. It is called the Wigner distribution. The Wigner distribution implies that the eigenvalues have mutual correlations and repel each other. Moreover, where the eigenvalues have no correlation and are randomly distributed, $P(s)$ becomes

\begin{equation}
P(s) = \exp ({-s}),
\label{eq4}
\end{equation}

this is known as the Poisson distribution in $RMT$. Since in real systems the matrix size is finite, the method of rescaling of the eigenvalues called unfolding has to be applied. For details about the unfolding method, see \cite{Kikkawa-1}. Additionally, we tested the $P(s)$ obtained by the one-sample Kolmogorov-Smirnov test. Also, the significance level $\alpha$ is set at $0.05$. When the p-value is larger than the significance level  $\alpha$, the null hypothesis is not rejected.

{\bf Inverse participation ratio $(IPR)$.} We used the inverse participation ratio $(IPR)$ to analyze the localization properties of the eigenvectors, which has been introduced by Bell and Dean, and initially was applied in the context of atomic physics \cite{Bell}. Specifically, $IPR$ is one of the most frequently used techniques in $RMT$ to identify nodes appearing in the localized eigenvectors. Throughout this study, we have calculated the eigenvalues and eigenvectors to determine the $IPR$ which tries to provide a measure for participation rates of all nodes in each eigenvalue. $IPR$ of each eigenvalue $\lambda_k$ is given by

\begin{equation}
IPR_{k}=\Sigma_{n=1}^{N}(u_{n}^{k})^4,
\label{eq5}
\end{equation}

where the $u_n^k$ stands for components of the related eigenvector, and $N$ is the number of nodes. The summation is over all elements in each eigenvector. The $IPR_{k}$ shows two extreme cases: (1) a vector with equal elements $(u_n^k = 1/\sqrt N) $ has $IPR_{k}=1/N$, (2) a vector with zero elements except one element $(u_n^k = 1)$ has $IPR_{k}=1$. Thus, the $IPR$ expresses the reciprocal of the number of eigenvector elements that contribute remarkably \cite{Jalan2}. Note that $IPR$ defined as above separates the top contributing genes by keeping the threshold. The threshold is taken via the following steps. First, we shuffle the network for $100$ iterations. Then, the eigenvalues that are out of the bulk of the shuffled networks are selected. Finally, the mean value of $IPR$ of shuffled versions plus three standard deviations is taken as the boundary. We additionally calculate the average $IPR$ to measure an overall localization of the network as

\begin{equation}
\langle IPR \rangle=\Sigma_{k=1}^{N} IPR_{k}.
\label{eq6}
\end{equation}

{\bf Node participation ratio $(NPR)$.} Despite $IPR$, which deals with each eigenvector, the node participation ratio $(NPR)$ is a characteristic of each node. To calculate the $NPR$ of each gene, we calculate the sum of its contributions in all eigenvectors as

\begin{equation}
NPR_{n}=\Sigma_{k=1}^N(u_n^k)^4.
\label{eq7}
\end{equation}

The summation is over the participation of a node in all eigenvalues. If a node has a random performance in all eigenvalues, then its $NPR$ is small, that is in extreme cases equal to $1/N$. If, however, a node has unique roles in a few numbers of eigenvectors, then the value of its $NPR$ is high, i.e., at extreme it equals to one. Thus, the great value of $NPR$ means that a node plays a critical role in a specific biological process. To identify genes with high $NPR$ among those with top contributions in the localized eigenvectors, we performed the same procedure of taking the threshold for the $IPR$ as mentioned above.

\section*{Results}

{\bf Structural properties of the networks.} The structural parameters of the networks, which are in line with our main research question, are presented in Table \ref{table1a}. Besides, the clustering coefficient of each node $CC_{i}$, which is calculated through Eq \eqref{eq1}, the clustering coefficient of the network $CC$ is achieved through Eq \eqref{eq2}. The point is that the higher value of $CC$, the more presence of functional modules in the network \cite{Barabasiii}. In other words, the higher value of $CC$ means the presence of a high number of clique structures in a network \cite{Watts}, which as building blocks of a network make it more robust \cite{Alon} and stable \cite{Dwivedi}. What follows is that the nonessential network has fewer cliques of order three compared to the essential one.  This indicates that there is a deficiency of building blocks in the nonessential network, which may be leading to a more unstable system. It should be remarked that besides the aforementioned reason about the significance of $CC$, it is a deterministic indicator in defining the network's topology. Therefore, to obtain a deeper insight into our networks' topology, which unifies the mentioned results above, we examine our networks' structure within the analytical expression for $CC$ of theoretical networks. The values of analytical expressions for $CC$ of theoretical networks, and that of our real data are presented in Table \ref{table1b}. The result indicates that the value of $CC$ in our networks, as in the small-world networks, is between the values of that in the random and lattice versions.

\newcolumntype{P}[1]{>{\centering\arraybackslash}p{#1}}
\newcolumntype{M}[1]{>{\centering\arraybackslash}m{#1}}
\begin{table}[h]
	\caption{{\bf Network analysis.}}
	\subfloat[Network parameters.]{
	\label{table1a}
	\centering
	\setlength{\extrarowheight}{3pt}
	\begin{tabular}{|c|c|c|c|c}
		\rowcolor{lightgray}
		\hline
		&  $\bf{Essential}$ & $\bf{Nonessential}$  \\
		\hline
		$\bf{Nodes}$  &  $1,040$   & $4,430$  \\
		\hline
		$\bf{Density \ of \ links}$  &  $0.006$ & $0.005$  \\ 	
		\hline

		$\bf{\langle k \rangle}$ &  $6.366$   & $24.207$ \\
		\hline
		$\bf{k_{max}}$ &  $28.120$   & $78.333$  \\
		\hline

		$\bf{Assortativity}$& $-0.160$  & $-0.139$ \\ 
		\hline
		$\bf{Clustering \ Coefficient}$  & $0.025$   & $0.007$  \\ 
		\hline

		$\bf{\langle IPR \rangle}$& $0.0031$     & $0.0007$  \\
		\hline
		$\bf{\langle NPR \rangle}_{imp \ genes}$& $0.0040$     & $0.0008$  \\
		\hline
		$\bf{\lambda_{max}}$ & $37.25$    & $39.29$  \\
		\hline

	\end{tabular}
}	\hspace{1cm}
	\subfloat[Clustering Coefficient $(CC)$ for canonical topologies.]{
	\label{table1b}
	\begin{tabular}{|c|c|c|c|c|c}
		\rowcolor{lightgray}
		\hline
		$\bf{Network \  type}$  &  $\bf{Expression}$ &$\bf{Essential}$ & $\bf{Nonessential}$ \\
		\hline
		$\bf{Lattice}$     &  $\frac{3(k-2)}{4(k-1)}$ &$CC_{ess} < 0.610$ & $CC_{non} < 0.718$ \\
		\hline
		$\bf{Random}$  &   $\frac{k}{N}$& $0.006 < CC_{ess} $ & $0.005 < CC_{non}$ \\
		\hline
	
		$\bf{Small-world}$   &   $\frac{3(k-2)}{4(k-1)}(1-p)^3$& $CC_{R}<0.445<CC_{L}$ & $CC_{R}<0.523<CC_{L}$ \\	
		\hline	

	\end{tabular}
	\hspace{.5cm}}
	\caption*{(a) Number of nodes, mean degree, maximum degree, density of links, assortativity, clustering coefficient, mean IPR, mean NPR of important genes, and maximum eigenvalue in both essential and nonessential gene networks are presented.}
	\caption*{(b) $CC_{ess} = 0.025$, $CC_{non} = 0.007$, $CC_{L}$, and $CC_{R}$ indicate the clustering coefficient of essential, nonessential, lattice, and random networks, respectively. The value of $CC_{ess}$ and $CC_{non}$ are between the values of that in random and lattice versions as small-world networks. $k$ = mean degree; $N$ = number of nodes; $p$ = the probability of rewiring in the small-world network, which is 0.1 here. Expressions have been adapted from \cite{van}.}

\end{table}

Additionally, the essential and nonessential gene networks exhibit approximately similar statistics for the assortativity coefficient. That is, as exhibited by most of the biological systems investigated under the network theory framework, our gene networks are disassortative \cite{Shinde}.  Specifically, components with high degrees tend to be connected with different ones. Moreover, the density of networks by summing over absolute magnitude values of links divided by all possible links shows high sparsity in our networks. Besides the analyzed structural features, the vital differences between them, are revealed through spectral analysis. In the following, the global spectral properties, eigenvalue fluctuations, and properties of genes appearing in the localized eigenvectors are analyzed. These are known as the most frequently used techniques in $RMT$ to understand the underlying complex system comprehensively.

\begin{figure}[!h]{
		(A)\includegraphics[height=5.cm,width=.5\linewidth]{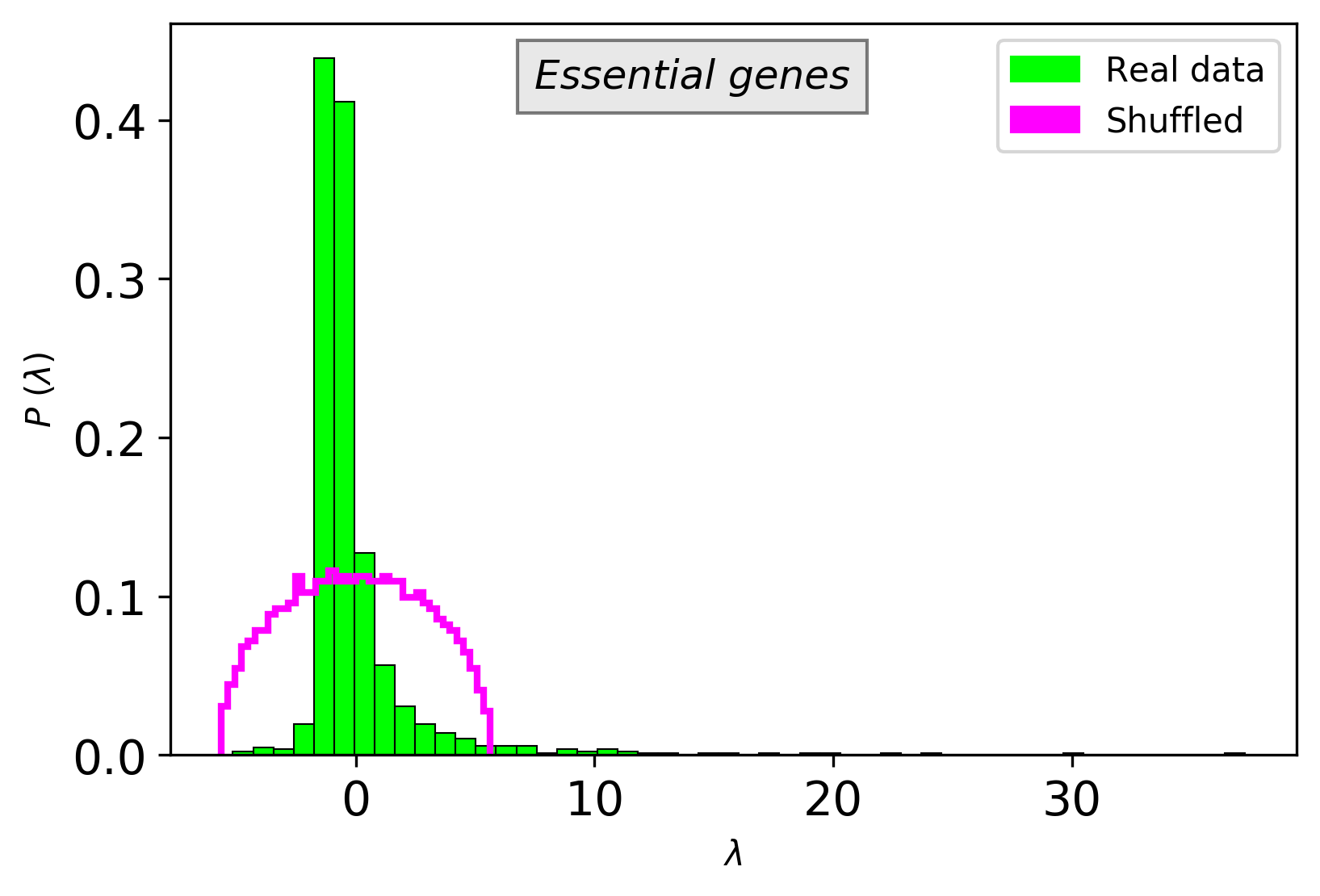}
		(B)\includegraphics[height=5.cm,width=.5\linewidth]{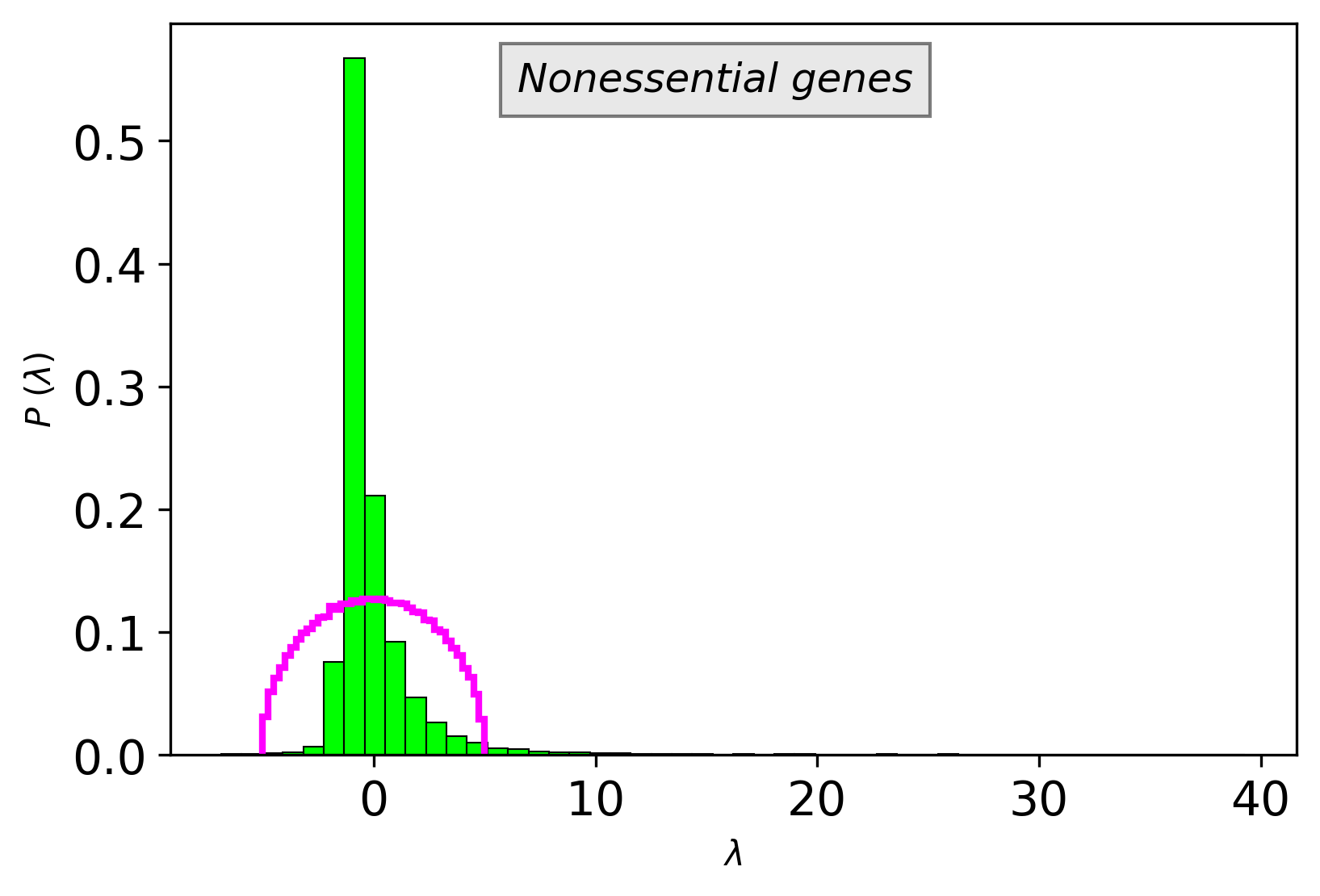}\\
		(C)\includegraphics[height=5.cm,width=.5\linewidth]{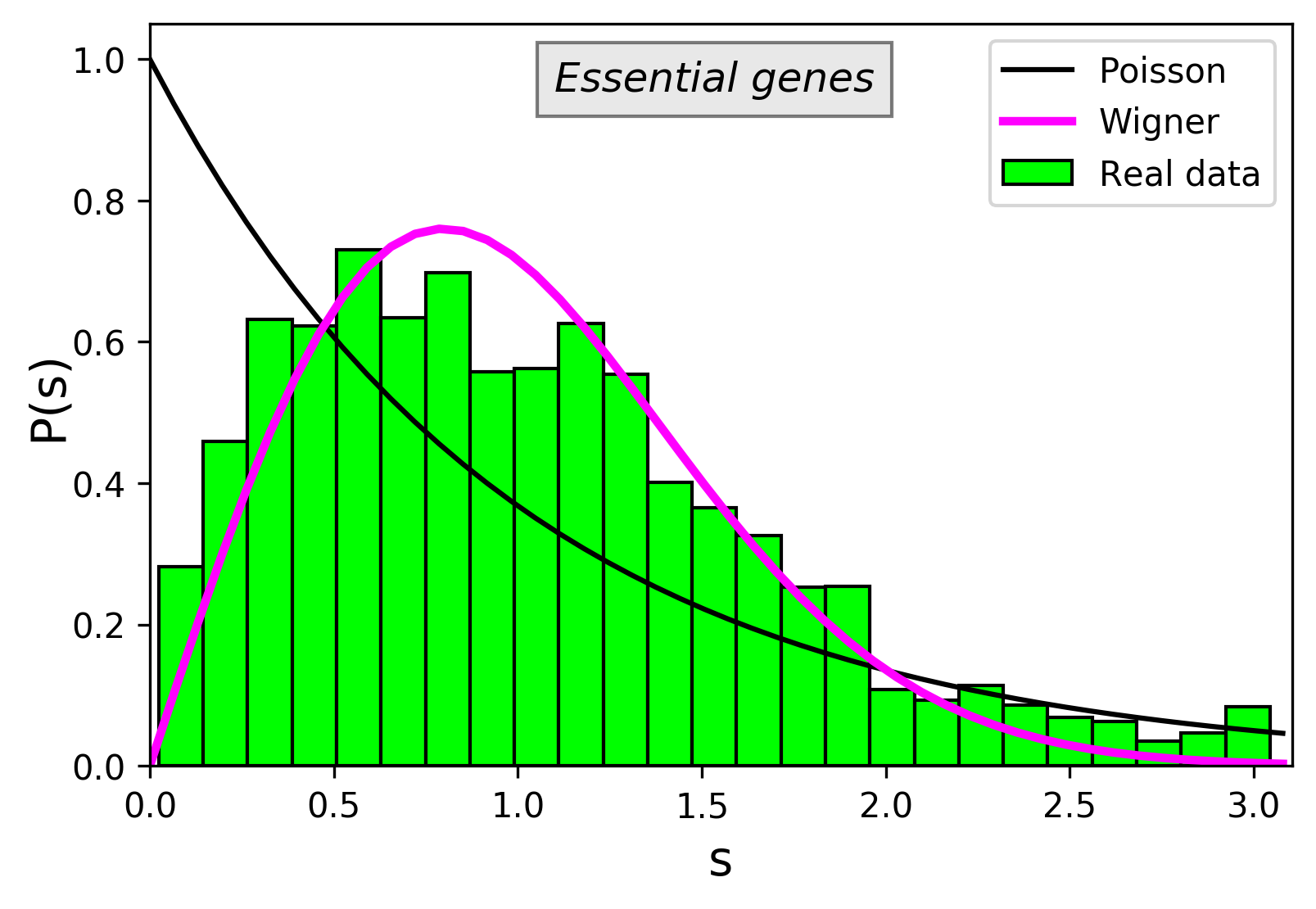}
		(D)\includegraphics[height=5.cm,width=.5\linewidth]{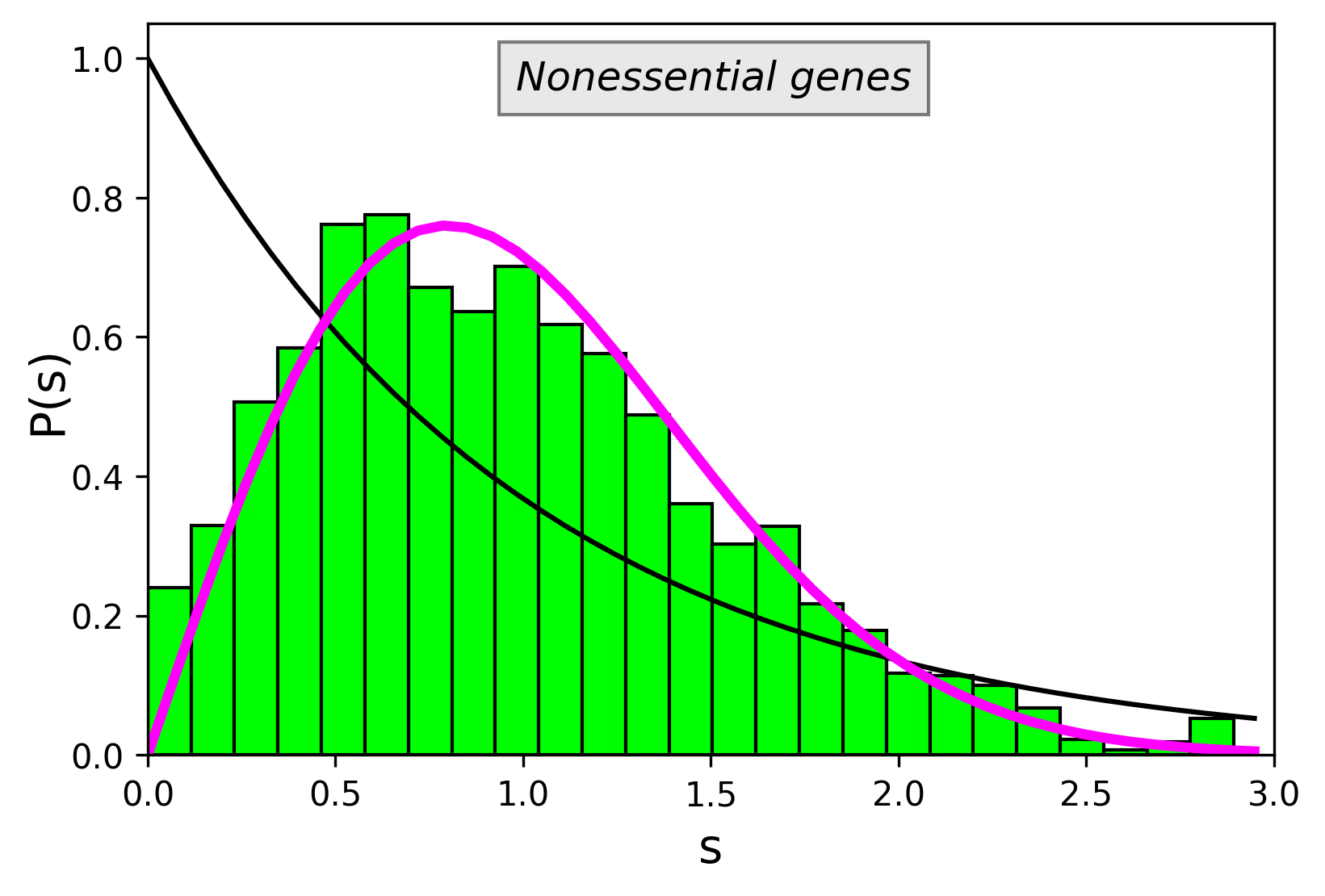}
	}
	\caption{{\bf Eigenvalue distribution and nearest neighbour spacing distribution for both essential and nonessential gene networks.} (A) $P(\lambda)$ for the essential gene network, (B) $P(\lambda)$ for the nonessential gene network. (The plots depict a clear difference between distribution for both networks (in green) and their corresponding shuffled versions (in purple).)(C) $P(s)$ for the essential gene network, (D) $P(s)$ for the nonessential gene network.  (The spacing of unfolded eigenvalues follows $GOE$ statistics of the Wigner distribution. The green bars represent data points, and the solid purple (black) line represents the Wigner (Poisson) distribution. Besides, the one-sample Kolmogorov-Smirnov test in each of the segments is applied which shows the p-values larger than the significance level $\alpha = 0.05$.)}
	\label{fig2}
\end{figure}

{\bf Eigenvalue distribution $P(\lambda)$ and nearest neighbour spacing distribution $P(s)$.} Now, a couple of questions arise: How do we evaluate the heterogeneities of the structures \cite{Cimini, Moradimanesh, Kargaran}? What happens if we analyze the nonessential gene interaction network? Are we left with a homogenous network with a random structure, then? To answer these questions, we can use spectral methods to look over the spectrum of eigenvalues \cite{Rai-3, Chan}. Indeed, if the pairwise interaction of genes does not have a higher level of structure, then its matrix should have properties of random matrices. From $RMT$, we know that for a system that consists of a large number of elements, if the structure is limited to the pairwise level without a higher scale structure and if the probability density function of the pairwise interactions comes from a Gaussian distribution, then the $P(\lambda)$ has a semi-circle shape that centers around zero. As the result, we end up with a distinct $P(\lambda)$ of the genetic interaction networks (in green bars) and their corresponding shuffled versions (in a nice purple semi-circle) in Fig\ref{fig2}A and Fig\ref{fig2}B. Despite the shuffled networks, the genetic interaction networks possess several eigenvalues that have large values which their related eigenvectors can carry meaningful information concerning the system \cite{Safavi}. The serious difference between the spectrum of the genetic interaction networks and their shuffled networks leads us to a clear conclusion that the genetic interaction networks have a structure with their communities and are far from randomness. Thus, significant information should be carried by the structure of the genetic interaction networks beyond the local pairwise interactions. About the second question, we know that the role of essential genes is so important in the network that their sole annihilation leaves a phenotype footprint in the reproduction process. However, we are surprised that nonessential genes whose sole role is not so important compared to essential ones, which have a vital phenotype footprint, play a role in shaping the heterogeneities in their network from the $RMT$ perspective.

Further, as depicted in Fig\ref{fig2}C and Fig\ref{fig2}D, both the essential and nonessential gene networks follow Gaussian orthogonal ensemble $(GOE)$ statistics of $RMT$. The distribution of spacing of unfolded consecutive eigenvalues follows the Wigner distribution (Eq \eqref{eq3}), not the Poisson distribution (Eq \eqref{eq4}). This fitness reflects that there is a correlation between eigenvalues. Thus, they repel each other; otherwise, there was no correlation between them. Therefore, it can be noted that both networks have a minimal amount of randomness which is referred to as random connections between nodes \cite{Jalannnn}. The universal $GOE$ statistics displayed by both networks lead to two main results. First, both networks, even the nonessential network have robustness. Second, it establishes that these networks can be modeled using the $GOE$ of $RMT$. Therefore, we can apply all the techniques developed under the well-established framework of $RMT$. In the following, by analyzing the localization properties of eigenvectors, the set of genes that have the most contribution to the structure in each network is achieved.

\begin{figure}[h]{ 
		(A)\includegraphics[height=5.cm,width=.5\linewidth]{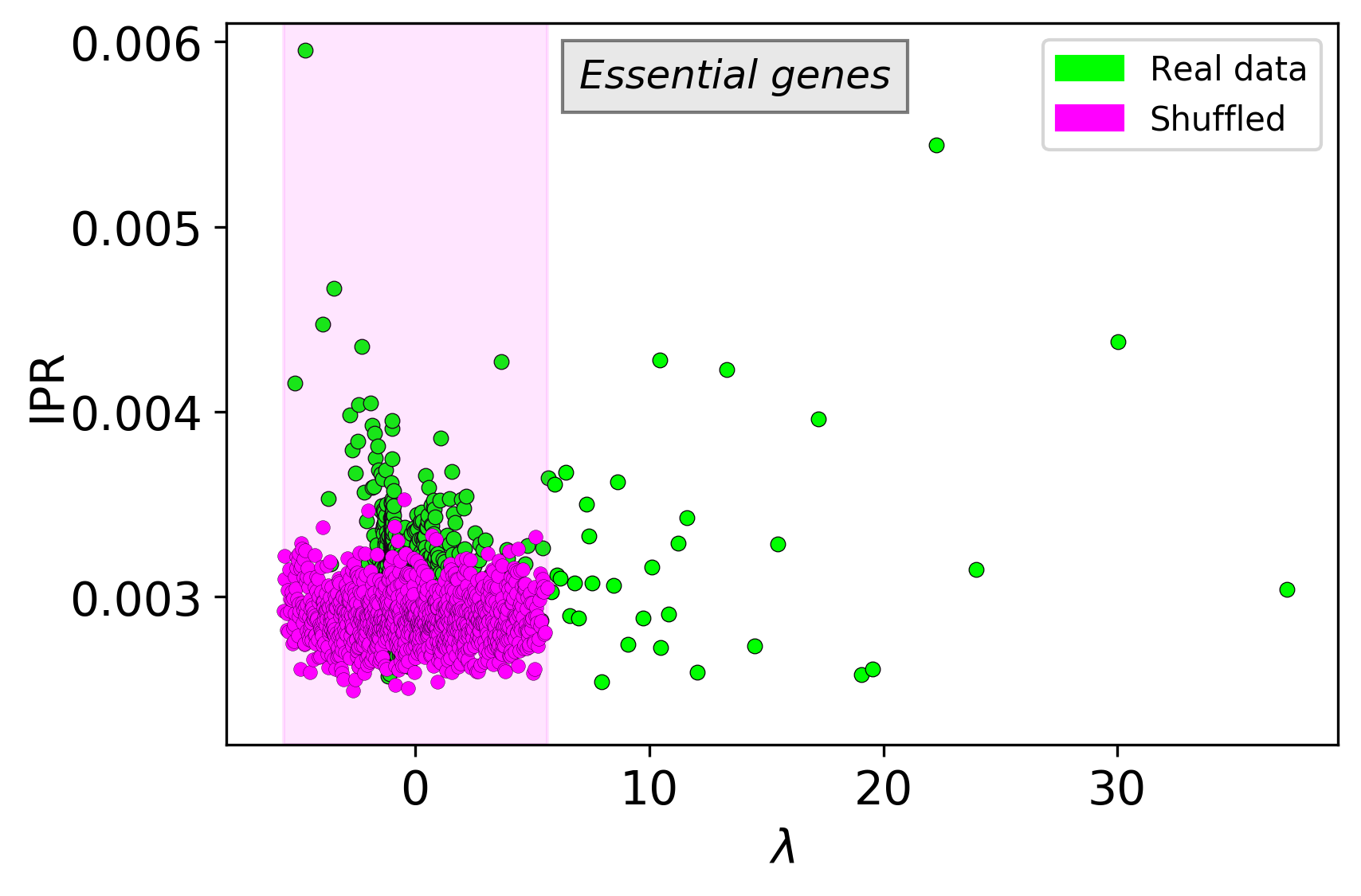}
		(B)\includegraphics[height=5.cm,width=.5\linewidth]{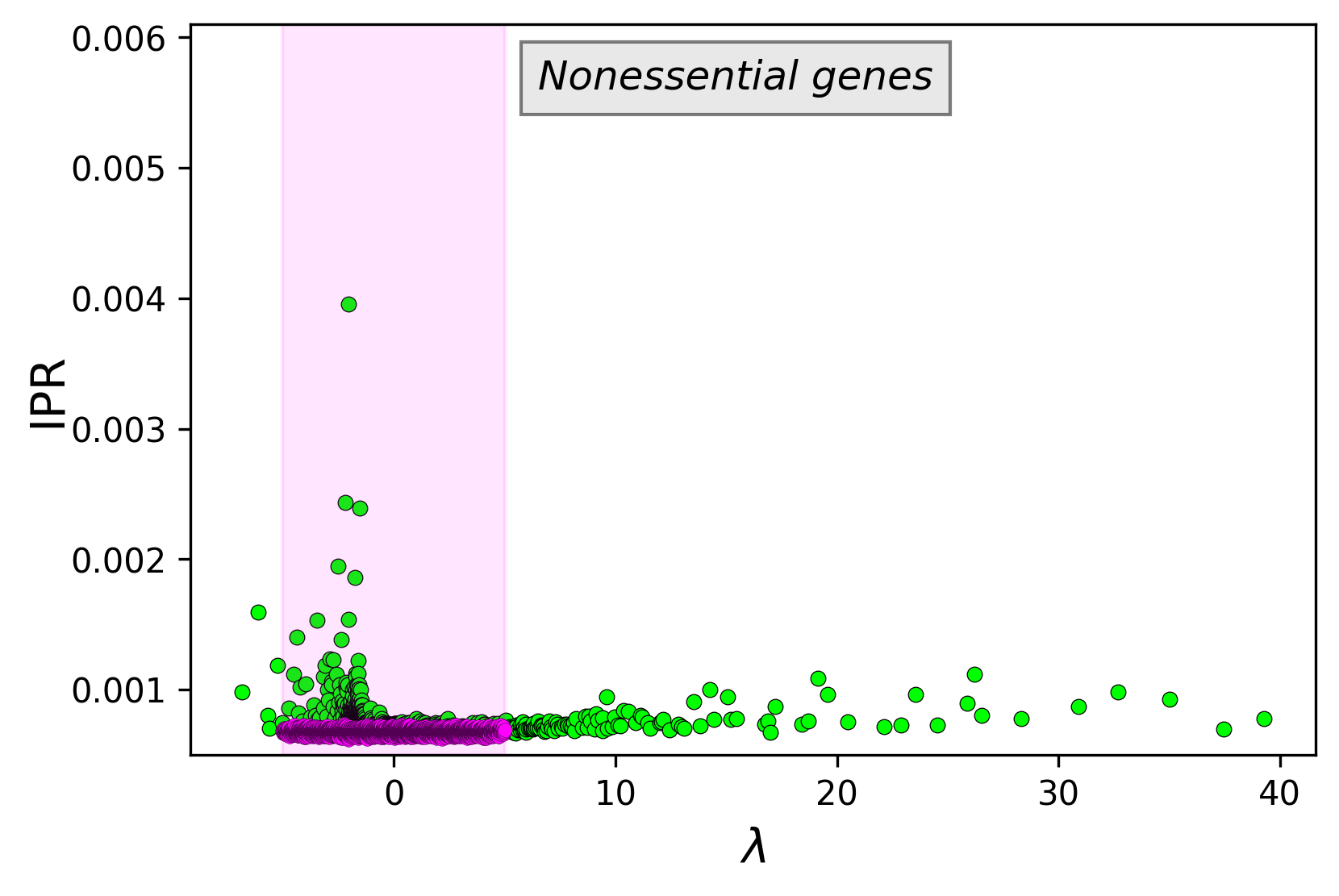}	
		(C)\includegraphics[height=5.cm,width=.5\linewidth]{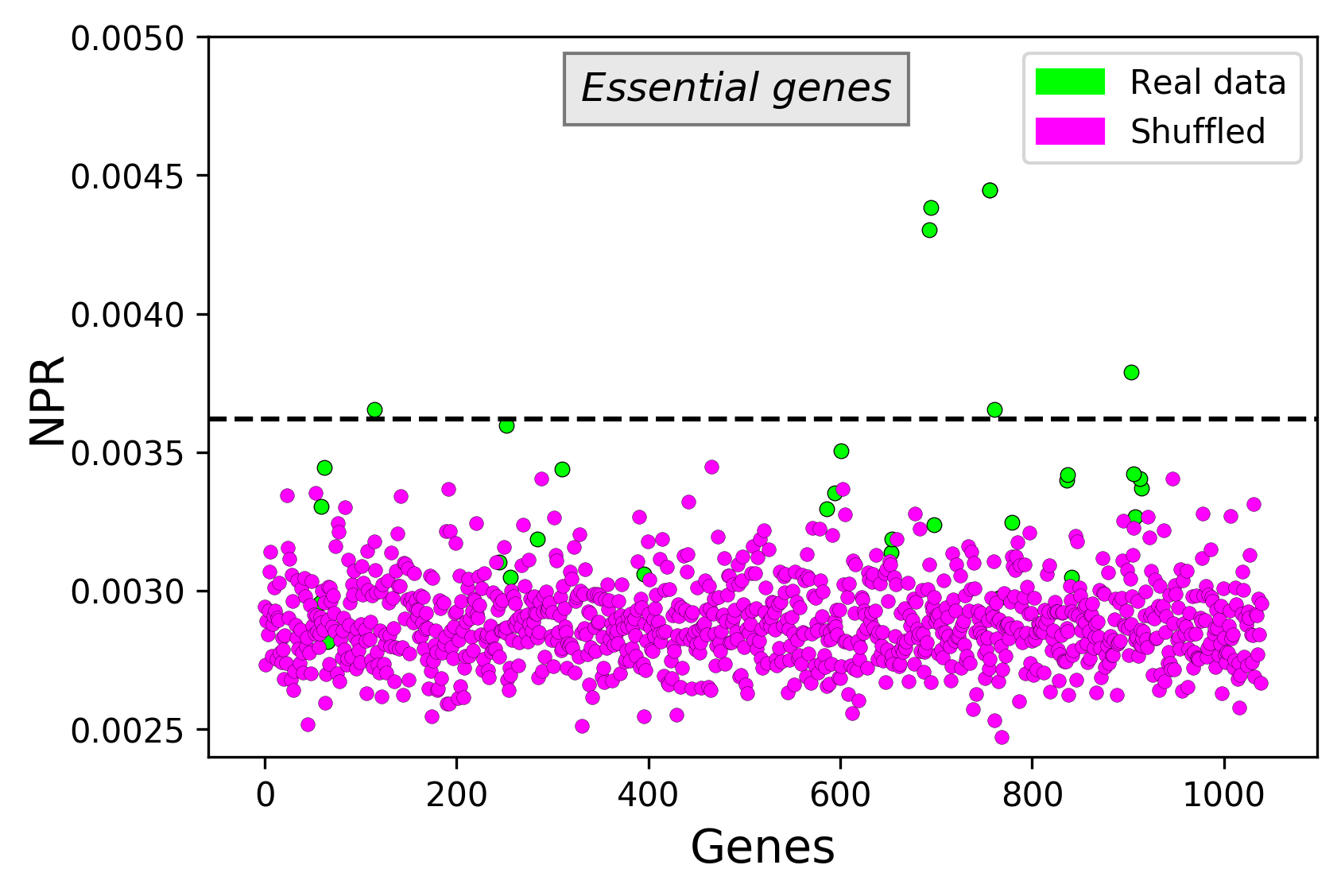}
		(D)\includegraphics[height=5.cm,width=.5\linewidth]{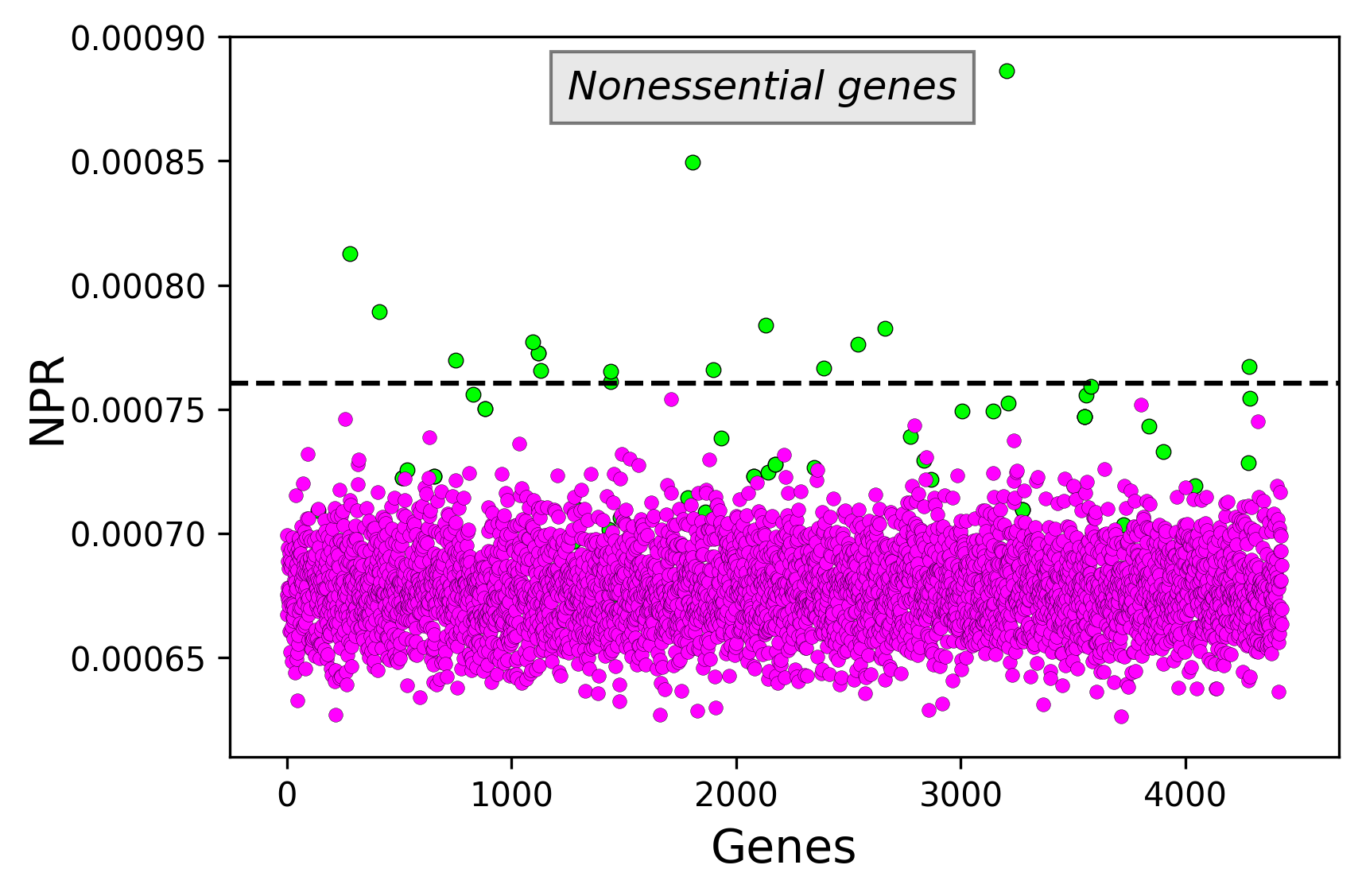}

	}
	\caption{{\bf Inverse participation ratio $(IPR)$ of eigenvalues and node participation ratio $(NPR)$ of genes  for both essential and nonessential gene networks.} (A) $IPR$ for the essential gene network, (B) $IPR$ for the nonessential gene network. (C) $NPR$ for the essential gene network, (D) $NPR$ for the nonessential gene network. (The green dots represent $IPR(NPR)$ of the real data, and the purple dots indicates that of shuffled versions.)}
	\label{fig4}
\end{figure}

{\bf Important genes through eigenvector localization.} Based on the inverse participation ratio $(IPR)$ values calculated using Eq \eqref{eq5}, the eigenvectors can be categorized into two groups. Some of them follow $RMT$ predictions of Porter-Thomas distribution which corresponds to random interactions \cite{Zyczkowski}. The other ones deviate from this universality and show localization which gives the system-dependent information  \cite{Plerou}. The deviating part reveals important genes as explained below. The essential (nonessential) gene network yields 26 (52) top contributing genes corresponding to the top 7 (11) most localized eigenvectors. The $IPR$ of the eigenvalues of the networks compared to their corresponding shuffled versions have been depicted in Fig\ref{fig4}A and Fig\ref{fig4}B. Moreover, the average $IPR$ calculated by using Eq \eqref{eq6} is higher for the essential network than for the nonessential one. It demonstrates that the nonessential network is more random than the essential network. In the following section, the functions of these selected genes are discussed briefly.

{\bf Functional properties of top contributing genes in the most localized eigenvectors.} The most important outcome of the functional analysis of top contributing essential genes in the most localized eigenvectors is that most of them are involved in the specific bioprocesses which particularly exist in the essential gene network (first colomn in Table \ref{table0}). The first localized eigenvector has five contributing genes, of which $YIL118W$ and $YLR166C$ are responsible for \enquote{cell polarity and morphogenesis} which occurs only in the essential gene network, $YGL116W$ in the cell controls the \enquote{mitosis and chromosome segregation}, and $YLR129W$, $YER127W$ help in \enquote{rRNA and ncRNA processing} in the essential network. Besides, the second localized eigenvector consists of four top contributing genes, among which $YFL005W$ and $YGL233W$ and $YLR166C$ as above are responsible for \enquote{cell polarity and morphogenesis}, and $YER136W$ is responsible for \enquote{glycosylation, protein folding/targeting, cell wall biosynthesis}. Also, among the top contributing genes in the third most localized eigenvector, $YKR086W$ is generally involved in \enquote{mRNA and tRNA processing}, $YNL163C$ is implicated in \enquote{ribosome biogenesis}. Similarly, $YPR103W$ controls \enquote{protein degradation/turnover}, and $YMR005W$, is found to handle \enquote{transcription and chromatin organization}. Moreover, the fourth localized eigenvector has two top contributing genes of which $YBR198C$ and $YDR145W$  are responsible for \enquote{transcription and chromatin organization}. What follows is that all the top contributing genes, in the fifth most localized eigenvectors, i.e., $YIR010W$ and $YMR117C$ and $YGL093W$ have a major contribution to \enquote{mitosis and chromosome segregation}. Among the top contributing genes of the sixth localized eigenvector, $YDR498C$ and $YLR078C$ have been found in \enquote{vesicle traffic}, $YHR036W$ accounts for \enquote{nuclear-cytoplasmic transport}, besides \enquote{mitosis and chromosome segregation}, and $YBL097W$ appear in \enquote{mitosis and chromosome segregation}. At last, all top contributing genes in the seventh localized eigenvalue contribute in \enquote{protein degradation/turnover}.

Regarding the functional analysis of the top contributing nonessential genes in the most localized eigenvectors, all of them interestingly are responsible for \enquote{respiration, oxidative phosphorylation, mitochondrial targeting}, bioprocess, which in particular exist in the nonessential gene network. Besides, most of the nonessential genes in the localized eigenvectors (33 of 52) take participate in one of the other specific bioprocesses in the nonessential gene network, which is \enquote{ribosome biogenesis}. At the third stage, 28 genes are responsible for \enquote{transcription and chromatin organization}, and \enquote{mitosis and chromosome segregation} which are common processes between essential and nonessential gene networks. What follows is the consecutive processes that have less number of responsible important genes, such as \enquote{vesicle traffic} (24 genes), \enquote{glycosylation, protein folding/targeting, cell wall biosynthesis} (19 genes), \enquote{DNA replication and repair} (17 genes). Also, \enquote{peroxisome} as a specific process in the nonessential gene network is devoted to 16 genes. Moreover, \enquote{metabolism and fatty acid biosynthesis}, as another specific process is handled through 8 important genes in the nonessential gene network. The three remained bioprocesses which are \enquote{tRNA wobble modification} and \enquote{MVB sorting and pH-dependent signaling}, and \enquote{mRNA processing} are driven with 6, 5, and 2 important genes in the nonessential gene network, respectively. Also, on average, most of the genes in each eigenvector are helping to four processes, that half of them are specific for the nonessential gene network, and the other half of them are common between essential and nonessential gene networks.

{\bf Node participation ratio $(NPR)$}. As the last step, to indicate genes that have played a significant role in a small portion of eigenvectors among those who live in the structure, we calculate the node participation ratio $(NPR)$ of each important gene through Eq \eqref{eq7}. Indeed, $NPR$ designates to identify if a gene has randomly taken part in a large number of eigenvectors or it has selected relatively a small portion of eigenvectors to take part. We calculate the $NPR$ of important genes achieved from $IPR$ analysis and depict the spectrums in Fig\ref{fig4}C and Fig\ref{fig4}D. As can be seen, the values of $NPR$ of essential genes are higher than those of nonessential genes. This means that essential genes compared to nonessential genes are strongly managing their network's structure. Moreover, this implies that besides a global contribution among those significant genes in nonessential genes, they weakly affect their structure. Particularly, 5 (15) specific essential (nonessential) genes indicate higher values of $NPR$ than the boundary of the shuffled version, which is achieved via the mean value of  NPR in the shuffled version after 100 iterations then added to three standard deviations.

{\bf Structural properties of localized genes with high value of $NPR$.} The top contributing genes in the most localized eigenvectors with high values of $NPR$ reveal interesting structural properties. Specifically, most of them in the essential gene network lie in the low degree regime compared to the mean degree, remarkably. But the importance of these genes based on the analysis of their interactions via $CC$ strikingly exhibits the existence of robust local structural patterns against external perturbations. Unlike the significant essential genes, the significant nonessential genes have higher degrees than their mean degree on average. Though, they still indicate a robust structure within the clustering coefficient higher than the clustering coefficient of their network. Besides that their gene ontology has been checked through the data file $S5$ in the dataset. Interestingly these important genes imply information on handling more purely specific bioprocesses which are merely appointed to the same network (first and second columns in Table \ref{table0}). All these significant genes, the specific bioprocesses handled by them, the values of their degrees, and the values of their clustering coefficients in both essential and nonessential networks are presented in Table \ref{table2a} and  Table\ref{table2b}, respectively.

\newcolumntype{P}[1]{>{\centering\arraybackslash}p{#1}}
\newcolumntype{M}[1]{>{\centering\arraybackslash}m{#1}}
\begin{table}[!h]
\caption{{\bf Classification of important genes with high NPR values.}}
\subfloat[Essential genes.]{
	\label{table2a}
	\centering
	\fontsize{5}{11}
	\setlength{\extrarowheight}{3pt}
	\begin{tabular}{|c|c|c|c|c}
	\rowcolor{lightgray}
	\hline
$\bf{Gene}$ &   $\bf{Biological \ process}$  & $\bf{Degree}$ & $\bf{Clustering \ Coefficient}$  \\
\hline
$YGL116W$ & \shortstack{ \\$protein \ degradation$ \\ $mitosis \ and \ chromosome \ segregation$} & $4.5$ & $0.073$ \\
\hline		
$YLR166C$ & \shortstack{ \\$cell \ polarity \ and \ morphogenesis$\\$protein \ degradation,$ $vesicle \ traffic$} & $5.7$ & $0.040$ \\
\hline
$YGL233W$ & \shortstack{ \\$cell \ polarity \ and \ morphogenesis,$\\$vesicle \ traffic$} & $7.0$ & $0.042$ \\
\hline
$YDR145W$ & \shortstack{ \\$cell \ polarity \ and \ morphogenesis$\\$protein \ degradation$\\$transcription \ and \ chromatin \ organization$} & $3.4$ & $0.023$ \\
\hline
$YFR004W$ & \shortstack{ \\$protein \ degradation/turnover$ \\$cell \ polarity \ and \ morphogenesis$} & $1.5$ & $0.051$ \\
\hline
\end{tabular}
}	\hspace{1cm}
\subfloat[Nonesential genes.]{
\label{table2b}
\fontsize{5}{11}
\begin{tabular}{|c|c|c|c|c|c}
\rowcolor{lightgray}
\hline
$\bf{Gene}$ &   $\bf{Biological \ process}$  & $\bf{Degree}$ & $\bf{Clustering \ Coefficient}$  \\
\hline
$YEL018W$ & \shortstack{ \\$Nuclear-cytoplasmic \ Transport,$ \\$Mitosis \ and \ Chromosome \ Segregation,$ \\$Ribosome \ Biogenesis,$ \\$Respiration,$ $DNA \ Replication \ and \ Repair$} & $32.9$ & $0.009$ \\
\hline
$YGL134W$ & \shortstack{ \\$Peroxisome,$ $Glycosylation,$\\$Transcription \ and \ Chromatin \ Organization,$ \\$Respiration,$ $tRNA \ Wobble \ Modification,$ \\$Nuclear-cytoplasmic \ Transport,$ \\$Vesicle \ Traffic,$ $Ribosome \ Biogenesis$\\$MVB \ Sorting \ and \ pH-dependent \ Signaling$} & $45.1$ & $0.009$ \\
\hline
$YCL016C$ & \shortstack{ \\$Respiration,$ $DNA \ Replication \ and \ Repair,$ \\$Mitosis \ and \ Chromosome \ Segregation$} & $35.8$ & $0.008$ \\
\hline
$YJL030W$ & \shortstack{ \\$Mitosis \ and \ Chromosome \ Segregation,$ \\$Respiration,$ $Ribosome \ Biogenesis,$\\$Transcription \ and \ Chromatin \ Organization,$ \\$DNA \ Replication \ and \ Repair$} & $40.3$ & $0.010$ \\
\hline
$YER016W$ & \shortstack{ \\$Respiration,$ $Ribosome \ Biogenesis,$ $Vesicle \ Traffic$\\$Mitosis \ and \ Chromosome \ Segregation,$ \\$Transcription \ and \ Chromatin \ Organization$} & $36.5$ & $0.013$ \\
\hline
$YEL062W$ & \shortstack{ \\$Respiration,$ $Vesicle \ Traffic$\\$Mitosis \ and \ Chromosome \ Segregation,$  \\$Transcription \ and \ Chromatin \ Organization,$ \\ $Ribosome \ Biogenesis,$ $DNA \ Replication \ and \ Repair$} & $22.7$ & $0.013$ \\
\hline
$YDR095C$ & \shortstack{ \\$Respiration,$ \\$Transcription \ and \ Chromatin \ Organization
	$} & $3.9$ & $0.008$ \\
\hline
$YBR164C$ & \shortstack{ \\$Respiration,$ $Vesicle \ Traffic$  \\$Mitosis\  and \ Chromosome \ Segregation,$\\ $Ribosome \ Biogenesis,$ $DNA \ Replication \ and \ Repair$} & $23.4$ & $0.007$ \\
\hline
$YMR071C$ & \shortstack{ \\$Ribosome \ Biogenesis,$ \\$Mitosis\  and \ Chromosome \ Segregation,$ \\$Respiration,$ $Vesicle \ Traffic$} & $33.6$ & $0.014$ \\
\hline
$YHR012W$ & \shortstack{ \\$Respiration,$ $Vesicle \ Traffic,$ $Peroxisome,$ \\$Ribosome \ Biogenesis,$ $Glycosylation,$\\$Mitosis \ and \ Chromosome \ Segregation,$ \\$MVB \ Sorting \ and \ pH-dependent \ Signaling,$\\$Nuclear-cytoplasmic \ Transport,$ \\$tRNA \ Wobble \ Modification$ } & $24.1$ & $0.011$ \\
\hline
$YGL136C$ & \shortstack{ \\$Respiration,$ $tRNA \ Wobble \ Modification,$\\$Mitosis \ and \ Chromosome \ Segregation,$ \\$Ribosome \ Biogenesis,$ $Vesicle \ Traffic,$  \\$DNA \ Replication \ and \ Repair$} & $13.4$ & $0.010$ \\
\hline
$YHR141C$ & \shortstack{ \\$Ribosome \ Biogenesis,$ $Respiration,$ $Glycosylation,$ \\$Transcription \ and \ Chromatin \ Organization,$ \\$Metabolism \ and \ Fatty \ Acid \ Biosynthesis,$ \\$DNA \ Replication \ and \ Repair$} & $22.0$ & $0.011$ \\
\hline
$YLL027W$ & \shortstack{ \\$Respiration,$ $Glycosylation,$ \\$Transcription \ and \ Chromatin \ Organization,$ \\$Metabolism \ and \ Fatty \ Acid \ Biosynthesis,$ \\$Ribosome \ Biogenesis,$ $DNA \ Replication \ and Repair$} & $20.6$ & $0.013$ \\
\hline
$YPL275W$ & \shortstack{ \\$Respiration,$ $Ribosome \ Biogenesis$ \\$MVB \ Sorting \ and \ pH-dependent \ Signaling,$ \\$DNA \ Replication \ and \ Repair,$ \\$Metabolism \ and \ Fatty \ Acid \ Biosynthesis,$ \\$tRNA \ Wobble \ Modification$ } & $30.5$ & $0.014$ \\
\hline
$YJR159W$ & \shortstack{ \\$Respiration,$ $Glycosylation,$ \\$tRNA \ Wobble \ Modification,$ $Ribosome \ Biogenesis$} & $28.9$ & $0.015$ \\
\hline
\end{tabular}
\hspace{.5cm}}

\end{table}

\section*{DISCUSSION}

This study has analyzed the undirected, weighted, and signed networks of essential and nonessential gene interaction networks of Saccharomyces cerevisiae. It is worth mentioning that to consider the whole story of a gene interaction system, both the functional relation and the containing structural patterns include beneficiary information. Besides the importance of this structure, detecting the significant genes with top contributions in the most localized eigenvectors is the second step to be taken. Therefore, our study has shown the following results:

There is a rich literature in the network theory in order to make us well acquainted with the structure and to distinguish non-random patterns through measuring many structural parameters such as clustering coefficients, mean degree, assortativity, modularity, etc. While the essential gene network has less number of nodes with less mean degree, it has a more robust and stable structure due to the higher clustering coefficient. But it doesn't mean that the nonessential gene network doesn't benefit from structure since the clustering coefficient of nonessential gene network follows the same behavior of small-world structure as being between random and lattice measures.

This paper, to achieve a deeper understanding of the global structural patterns in our gene interaction networks, under the $RMT$ framework, focuses on the spectrum of eigenvalues. Specifically, the first major question in this paper is about the signs of structures, especially in the nonessential gene network. Our results interestingly suggest a clear difference between the spectrum of the eigenvalues of the interaction matrices for both essential and nonessential gene networks with their corresponding shuffled counterparts, which follow the semi-circle distribution of the random network. Moreover, the greatest eigenvalue in a larger (expectedly highly structural) network is bigger than that of a smaller one. While the greatest eigenvalue in the nonessential gene network has nearly the same value in the essential gene network. Thus, there is a weekly structure in the nonessential gene network, which takes it away from randomness. In addition, through the spectrum of eigenvalues, the fluctuation of them as a strong analysis in $RMT$ is analyzed. The results indicate that both essential and nonessential gene networks follow $GOE$ statistics of $RMT$. Thus, due to the universality, we are allowed to apply the $RMT$ method. On the other hand, this fitness with Wigner distribution reflects that both the networks have a minimal amount of randomness between their genes, which means there is a correlation between their eigenvalues so that they repel each other.

Another main goal is to identify genes that play a significant role in the structure of the networks. The genes that contribute to the heterogeneous structure are meaningful to us, and the value of their contribution is proportional to their eigenvector component. Inverse participation ratio $(IPR)$ is a measurement for the participation rates of all nodes in each eigenvalue. Regarding the $IPR$ of original networks, we have observed the following result: The average of $IPR$ in both networks is higher than that of shuffled ones. This result supports the fact that the networks have a substructure. Additionally, the higher value of $IPR$ in the essential gene network than the nonessential one indicates the existence of a stronger structure and more heterogeneities in this network. Furthermore, in Costanzo's study, besides all specific features of essential (nonessential) genes, they are responsible for three (seven) bioprocesses that are unique for themselves. In the analysis of the $IPR$, we observed that most of these top contributing genes in the most localized eigenvectors are responsible for those specific bioprocesses. Last but not least, this is the first time that node participation ratio $(NPR)$ is utilized to calculate the participation of each gene in all eigenvectors. A high value of $NPR$ identifies nodes that play a significant role in a small portion of the eigenvectors. Our result indicates that in both networks, the average value of $NPR$ of important genes is larger than that of corresponding shuffled versions. Moreover, essential genes have a larger value of $NPR$ than nonessential genes, which means that essential genes manage the structure strongly, while nonessential genes have a global role in their structure and weakly control it. Finally, in the essential (nonessential) gene network, we observed 5 (15) top contributing genes in the most localized eigenvectors with the highest value of $NPR$ that despite lying at a low (high) degree regime, exhibit structural significance with high clustering coefficients. This interesting result turns out to be more fascinating in the light of the functional analysis of these important genes, which mostly are responsible for the specific bioprocesses appointed to their network.

\section*{Author contributions}
{\bf Conceptualization:} Nastaran Allahyari, Gholam Reza Jafari, Ali Hosseyni, Nima Abedpour.
\\
{\bf Formal analysis:} Nastaran Allahyari, Gholam Reza Jafari, Ali Hosseyni, Nima Abedpour.
\\
{\bf Methodology:} Nastaran Allahyari, Gholam Reza Jafari, Ali Hosseyni, Nima Abedpour.
\\
{\bf Resources:} Nastaran Allahyari, Nima Abedpour, Gholam Reza Jafari.
\\
{\bf Software:} Nastaran Allahyari.
\\
{\bf Supervision:} Gholam Reza Jafari, Ali Hosseyni.
\\
{\bf Visualization:} Nastaran Allahyari.
\\
{\bf Writing:} Nastaran Allahyari.

\section*{Acknowledgments}
N.A. would like to express her appreciation to M. Ghanbarzadeh, A. Kargaran, and Z. Moradimanesh for constructive comments that improved the manuscript.

\end{document}